\documentclass[aps,prb,superscriptaddress,twocolumn]{revtex4-1}

\usepackage{amsmath, amsthm, amssymb}
\usepackage{graphicx}
\usepackage{color}
\usepackage{times}
\usepackage{dsfont}
\usepackage{epstopdf}

\usepackage{natbib}
\setcitestyle{numbers,square}

\usepackage{hyperref}
\hypersetup{
        colorlinks=true,
}

\DeclareMathOperator{\Tr}{Tr}
\newcommand{\id}{\mathds{1}}
\newcommand{\sem}{{\hspace{.4mm}\mathrm{s}}}

\newcommand{\eqnref}[1]{(\ref{#1})}

\newcommand{\ket}[1]{\mbox{$| #1 \rangle$}}
\newcommand{\bra}[1]{\mbox{$\langle #1 |$}}
\newcommand{\braket}[2]{\mbox{$\langle #1  | #2 \rangle$}}
\newcommand{\ingr}[2]{\begin{matrix}\includegraphics[height=#1 cm]{#2}\end{matrix}}
\newcommand{\pvec}[1]{\vec{#1}\mkern4mu\vphantom{#1}}

\begin{document}

\title{Semion wave function and energetics in a chiral spin liquid on the Kagome lattice}

\author{L. Cincio}
\affiliation{Perimeter Institute for Theoretical Physics, Waterloo, Ontario, N2L 2Y5, Canada}

\author{G. Vidal}
\affiliation{Perimeter Institute for Theoretical Physics, Waterloo, Ontario, N2L 2Y5, Canada}

\author{B. Bauer}
\affiliation{Station Q, Microsoft Research, Santa Barbara, CA 93106-6105, USA}

\begin{abstract}

Recent years have seen the discovery of a chiral spin liquid state -- a bosonic analogue of a
fractional Quantum Hall state first put forward by Kalmeyer and Laughlin in 1987 -- in several
deformations of the Heisenberg model on the Kagome lattice.
Here, we apply state-of-the-art numerical techniques to one such model, where breaking of the time-reversal
symmetry drives the system into the chiral phase. Our methods allow us to obtain explicit
matrix-product state representations of the low-lying excitations of the chiral spin liquid state, including the
topologically non-trivial semionic excitation. We characterize
these excitations and study their energetics as the model is tuned towards a topological phase transition.

\end{abstract}

\maketitle

\section{Introduction}

For decades, researchers have sought to find examples of spin liquid phases~\cite{balents2010}, i.e. Mott insulators
that are at lowest temperatures not characterized by local order parameters. While numerical simulations are
frequently invoked to explore these phases, the reliable characterization of spin liquid phases has remained a great challenge.
Recent years have seen a fast evolution of numerical methods that have facilitated a paradigm shift away from
characterizing spin liquid phases by the lack of local order towards a characterization through positive, universal signatures.
This is particularly the case for gapped, topologically ordered spin liquids, where an extensive theoretical framework
exists for the universal properties of these phases~\cite{nayak2008}.

A recent proposal to numerically extract the behavior under
modular transformations~\cite{zhang2012} has, where practically feasible, allowed an unambiguous
identification of the universal properties of a topological phase.
In Ref.~\onlinecite{cincio2012}, it was demonstrated that this technique can be applied to matrix-product state
approximations~\cite{white1992,white1992-1,ostlund1995} of the ground states of some system on thin,
infinitely long cylinders. This allows the extraction of both the entanglement spectrum~\cite{li2008} and the modular
$T$ and $S$ matrices that fully characterize topological bulk excitations using unbiased numerical
techniques for a large class of Hamiltonians.

Here we extend these methods and use them to construct explicit wavefunctions for isolated topologically
non-trivial and trivial excitations. We characterize these excitations through their SU(2) and momentum quantum
numbers and explore their energetics, allowing us to gain understanding of the excitation spectrum of
the topological phase.
We apply this method to an SU(2)-invariant spin-1/2 model on the Kagome lattice, where a recent
study~\cite{bauer2014} using the aforementioned techniques established the presence of a topologically
ordered chiral spin liquid phase~\cite{kalmeyer1987}. This phase emerges when time-reversal symmetry is broken in the
well-known Kagome Heisenberg antiferromagnet by adding a sufficiently strong coupling to the scalar spin
chirality~\cite{wen1989,baskaran1989}.

Beyond being able to gain access to explicit wavefunctions
for topological excitations in a spin system, our methods allow us to study the interplay of topological
order and SU(2) symmetry in this particular realization of the chiral spin liquid.
In a model that both has a microscopic symmetry and emergent topological order,
a possible scenario is that of symmetry fractionalization, where the anyonic excitations carry fractional
charge of the microscopic symmetry, as opposed to integral charge carried by conventional local excitations.
Famously, this is the case in the quantum Hall effect~\cite{tsui1982}, where the anyons carry fractional electric charge.
It is thus natural to expect that the excitations of the chiral
spin liquid, whose universal properties are described by a bosonic $\nu=1/2$ Laughlin state~\cite{kalmeyer1987},
carry fractional SU(2) charge~\cite{kalmeyer1989}. We will demonstrate that this is indeed the case in this
particular model.

Given our numerical characterization of the low-lying excitations,
we are able to establish a close connection between topological and non-topological degrees of freedom deep
in the chiral phase, where we show that the lowest spin excitations can be identified as pairs of non-interacting
fractional excitations. Tuning the model towards a topological phase transition out of the chiral phase, we observe
the onset of interactions between the fractional excitations. We furthermore observe a major reorganization
of the low-lying singlet excitations which appears to drive the transition out of the chiral phase.

The paper is structured as follows: In Section~\ref{sct:model}, we introduce the model that will be studied. In Section~\ref{sct:methods}, we describe an ansatz that allow us to obtain explicit wavefunction representations for excitations for infinite quasi-one-dimensional systems. In Section~\ref{sct:results}, we discuss our numerical results.

\section{Model}
\label{sct:model}

Following a number of exact parent Hamiltonians~\cite{yao2007,schroter2007,thomale2009,nielsen2013} as
well as realizations in topological flat-band models~\cite{tang2011,sun2011,neupert2011}, recent years have
seen the chiral spin liquid realized in a number of deformations of the Heisenberg model on the Kagome
lattice~\cite{bauer2014,he2014,gong2014}.
The model we focus on here is given by~\cite{bauer2014}
\begin{equation} \label{eqn:H}
H = \cos(\theta) \sum_{\langle i,j \rangle} \vec{S}_i \cdot \vec{S}_j + \sin(\theta) \sum_{\bigtriangleup} \vec{S}_i \cdot (\vec{S}_j \times \vec{S}_k ),
\end{equation}
where $\vec{S}$ denotes spin-1/2 operators acting on the sites of a Kagome lattice, and the sum in the first
term runs over all pairs of nearest neighbors, whereas the second sum runs over the elementary triangles of
the Kagome lattice and sites are ordered clockwise around each triangle.
The first term describes the well-known Heisenberg antiferromagnet, which has received much attention
over the years~\cite{elser1989,marston1991,sachdev1992}. Recent evidence from large-scale
density-matrix renormalization group calculations has favored a gapped phase without local order~\cite{yan2011}
which appears topologically ordered with a topological entanglement entropy of
$\log(2)$~\cite{jiang2012,depenbrock2012}; at the same time, calculations using variational
Monte Carlo hint at a gapless state~\cite{iqbal2011,clark2012}. Assuming the state is indeed gapped and topologically
ordered, the value of the topological entanglement entropy would be consistent with two distinct topological phases,
which we will refer to as "toric code"~\cite{kitaev2003} and "double semion"~\cite{freedman2004} phases, although their history as distinct $\mathbb{Z}_2$
gauge theories predates the toric code. While these phases cannot be distinguished
by their edge or their ground state degeneracy, their bulk excitations exhibit different exchange statistics. Which
of these phases is realized in a putative time-reversal symmetric topological phase in the Kagome Heisenberg antiferromagnet
is at this point an unresolved question; recent work has put forward energetically favorable ansatz states for both
candidates~\cite{poilblanc2013,qi2014,iqbal2014,buerschaper2014}, but has also placed severe symmetry
constraints on the double semion wavefunction~\cite{zaletel2014}.

The second term of Eqn.~\eqnref{eqn:H}, coupling to the scalar spin chirality~\cite{wen1989,baskaran1989}, breaks time-reversal
while preserving the SU(2) spin symmetry and leads to the formation of the chiral phase for
$\theta \gtrsim 0.05 \pi$~\cite{bauer2014}. The universal properties
of this chiral phase are captured by a bosonic $\nu=1/2$ Laughlin state, which exhibits a two-fold degeneracy
of its ground state on the torus or infinite cylinders, where the states can be characterized by different topological
flux through the torus or cylinder; bulk excitations that obey semionic statistics, i.e. the wavefunction
acquires a phase factor $i$ when two quasiparticles are interchanged; and a chiral gapless edge state described by an
SU(2)$_1$ Wess-Zumino-Witten (WZW) model with a chiral central charge of $c=1$.
This gapless edge is observable not only in the energy spectrum when placed on a system with a boundary, but
also in the entanglement spectrum~\cite{li2008}.
Other aspects of this phase have recently been studied using a variety of numerical~\cite{wietek2015} and
analytical~\cite{gorohovsky2015,meng2015} techniques.

No signs of time-reversal symmetry breaking
have been observed for the Heisenberg model itself. Therefore,
we expect a transition to take place between the chiral spin liquid phase and a putative time-reversal symmetric phase
surrounding the Heisenberg point. Such a transition has been analyzed recently in a different model in Ref.~\onlinecite{he2015},
as well theoretically in Ref.~\onlinecite{barkeshli2013}.
For the model considered here, we have shown~\cite{bauer2014} that the transition must take place
in the range $0 \leq \theta \leq 0.05 \pi$. Given the inherent limitations in terms of system size, which are accentuated
by the increasing correlation length as the phase transition is approached, it is challenging to locate the transition
more precisely and argue whether its nature is first-order or continuous.

The symmetry-enriched properties of this realization of the chiral spin liquid can be determined from the
entanglement spectrum reported in Ref.~\onlinecite{bauer2014}, where we found the degeneracies of the entanglement
spectrum to agree with the descendants of the two primary fields of the SU(2)$_1$ WZW model. Within this model, the states can be
organized into SU(2) multipletts~\footnote{See, e.g., Tables 15.1 and 15.2 in Ref.~\onlinecite{bigyellow}}.
In the tower of descendants of the identity field, the spins for the first four levels are $0$,$1$,$0 \oplus 1$,$0 \oplus 1 \oplus 1$;
we find this spectrum to be realized on the edge of a cylinder with trivial flux. Conversely, a cylinder with semionic flux realizes
the descendant spectrum of the spin-1/2 field, where we have $1/2$, $1/2$, $1/2 \oplus 3/2$, $1/2 \oplus 1/2 \oplus 3/2$.
We note that these SU(2) quantum numbers are emergent and the same level counting is found in models that lack a
microscopic SU(2) symmetry~\cite{tang2011,neupert2011,sun2011,bauer2014}. However, for models that do have
a microscpic SU(2) symmetry, we can assign to each level in the entanglement spectrum a micoscopic SU(2) quantum number.
Performing this identification for the ground states of~\eqnref{eqn:H}, we find that the microscopic quantum numbers
coincide with the emergent quantum numbers in the CFT spectrum, which implies that the semion carries a fractional spin of $S=1/2$.
We emphasize that no
local, topologically trivial excitation can carry such spin; a local spin flip, for example, corresponds to an $S=1$
excitation. Considering the fusion rule of the semion, $s \otimes s = \id$, and comparing this with the well-known
fusion rules for SU(2) spins, e.g. $\frac{1}{2} \otimes \frac{1}{2} = 0 \oplus 1$, we can generally identify
half-integer spin excitations as topologically non-trivial and integer spin excitations as trivial.

\section{An ansatz for localized excited states}
\label{sct:methods}

To access the properties of the model~\eqnref{eqn:H} computationally, we consider the Hamiltonian on a cylinder with infinite length and finite width. A key advantage of the geometry of infinite cylinders is that for a topologically ordered system, we can obtain the full set of degenerate ground states in a basis of states with well-defined topological flux \cite{cincio2012}.

\begin{figure}
  \includegraphics[width=0.99\columnwidth]{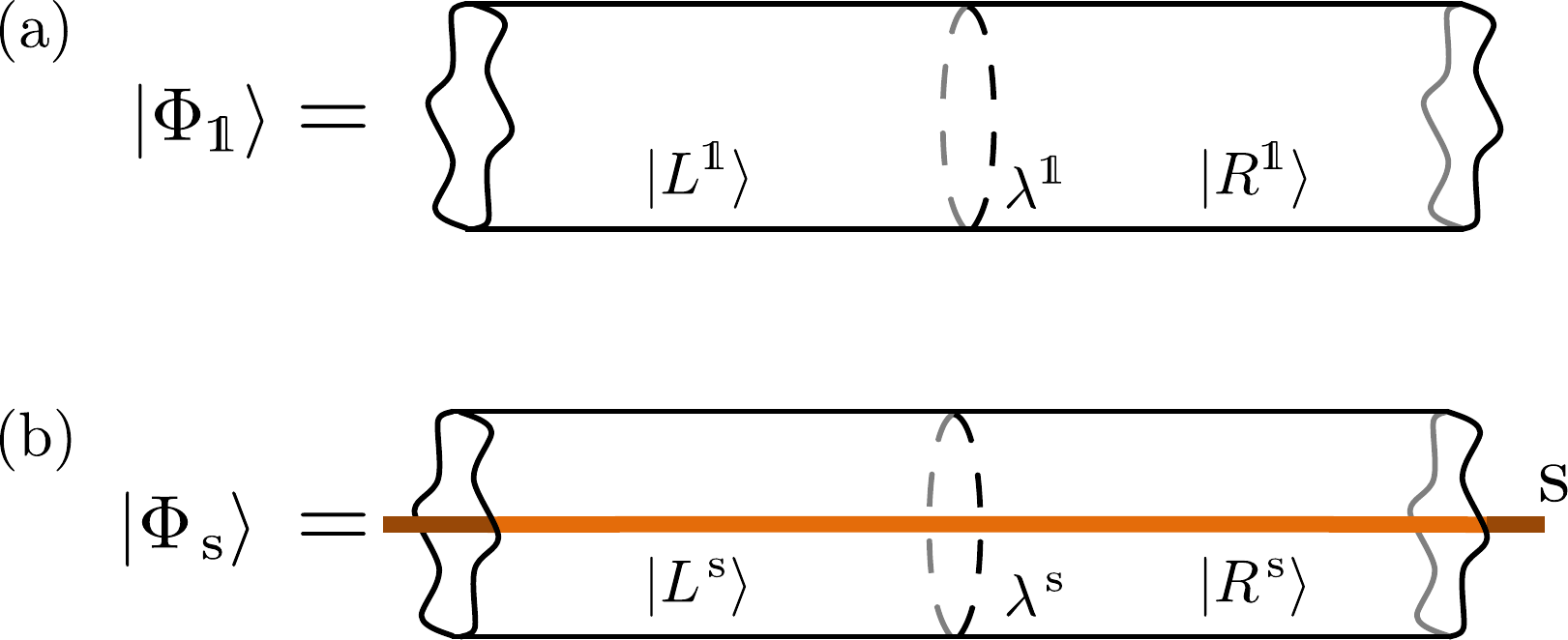}
  \caption{(Color online) The Schmidt decomposition of the ground states of the Hamiltonian (\ref{fig:gs}) on an infinite cylinder. $\ket{\Phi_\id}$ carries no topological flux while $\ket{\Phi_\sem}$ has a semion $\mathrm{s}$ propagating through the cylinder.
  \label{fig:gs} }
\end{figure}

For the Hamiltonian~\eqnref{eqn:H} on an infinite cylinder, restricted to the parameter range $0.05\pi \leq \theta \leq 0.5\pi$, Ref.~\onlinecite{bauer2014} finds two ground states: $\ket{\Phi_\id}$ and $\ket{\Phi_\sem}$. Here, $\ket{\Phi_\id}$ denotes a ground state with identity flux and $\ket{\Phi_\sem}$ has a semion threading through the cylinder. The Schmidt decomposition of $\ket{\Phi_\id}$ and $\ket{\Phi_\sem}$ with respect to partitioning the system into two semi-infinite cylinders reads
\begin{eqnarray}
\ket{\Phi_\id} &=& \sum_\alpha \lambda^\id_\alpha \ket{L^\id_\alpha} \ket{R^\id_\alpha} \ , \label{eqn:gs_id}\\
\ket{\Phi_\sem} &=& \sum_\alpha \lambda^\sem_\alpha \ket{L^\sem_\alpha} \ket{R^{\sem}_\alpha} \ \label{eqn:gs_s} ,
\end{eqnarray}
where $\ket{L_\alpha}$ ($\ket{R_\alpha}$) form an orthonormal basis of vectors of the left (right) partition of a cylinder. $\lambda_\alpha$ are the Schmidt coefficients and $\left(\lambda_\alpha\right)^2$ are the eigenvalues of the reduced density matrix of the left (right) semi-infinite cylinder. Fig.~\ref{fig:gs} gives a graphical representation of the Schmidt decomposition of the ground states $\ket{\Phi_\id}$ and $\ket{\Phi_\sem}$.

Our ansatz for the excited states is based on the idea of replacing the Schmidt coefficients $\lambda$ in (\ref{eqn:gs_id},\ref{eqn:gs_s}) by a matrix $G$ allowing for off-diagonal terms $\ket{L_\alpha}\ket{R_\beta}$ to contribute. The ansatz for the excitation takes the form
\begin{equation} \label{eqn:ansatz}
\ket{\Psi} = \sum_{\alpha,\beta} G_{\alpha\beta} \ket{L_\alpha} \ket{R_\beta} \ ,
\end{equation}
where the matrix $G$ is normalized such that $\braket{\Psi}{\Psi}~\equiv~\Tr (G^\dagger G) = 1$ and is energy
optimized under constraint $\braket{\Phi}{\Psi}~\equiv~\sum_\alpha \lambda_\alpha G_{\alpha\alpha} = 0$ to ensure that the excitation $\ket{\Psi}$ is orthogonal to the ground state $\ket{\Phi}$. The ansatz (\ref{eqn:ansatz}) represents an excitation that is localized in the longitudinal ($\hat{x}$) direction of a cylinder and completely delocalized in the transversal ($\hat{y}$) direction. Specifically, the excitation is localized within the correlation length in the $\hat{x}$ direction and has a has a well-defined transverse momentum $k_y$.
Throughout the remainder of this section, we will use the notation $\ket{\Psi^{a,\mu}}$ to denote different types of excitations, where
$a$ stands for the topological charge of an excitation, which can take the values $a=\id$ (identity)
and $\mathrm{s}$ (semion) for the case considered here, and $\mu$ labels its quantum numbers under the local symmetry.

For an SU(2) invariant ground state, an index $\alpha$ that labels the Schmidt vectors $\ket{L_\alpha}$, $\ket{R_\alpha}$ in (\ref{eqn:gs_id},\ref{eqn:gs_s}) carries SU(2) quantum numbers: the total spin $J$ and its z component $J_z$, where $\vec{J} \equiv \sum_i \vec{S}_i$ is the total spin operator. This feature allows us to access excitations with well-defined SU(2) quantum numbers by enforcing further constraints on the matrix $G$.
We will denote such excitations as $\ket{\Psi^{a,\mu}}$ with $\mu = S$ (singlet, $J=0$), $T$ (triplet, $J=1$) or $D$ (doublet $J=1/2$).

For example, a singlet excitation above the ground state with the identity flux is obtained by replacing the Schmidt coefficients $\lambda^\id$ in \eqnref{eqn:gs_id} by a matrix $G^{\id,S}$,
\begin{equation} \label{eqn:singlet}
\ket{\Psi^{\id,S}} = \sum_{\alpha,\beta} G^{\id,S}_{\alpha\beta} \ \ket{L^\id_\alpha} \ket{R^\id_\beta} \ ,
\end{equation}
where the only non-zero elements $G^{\id,S}_{\alpha\beta}$ are such that the indices $\alpha$ and $\beta$ fuse into a singlet. Matrix $G^{\id,S}$ in (\ref{eqn:singlet}) is the subject of an energy optimization.

\begin{figure}
  \includegraphics[width=0.99\columnwidth]{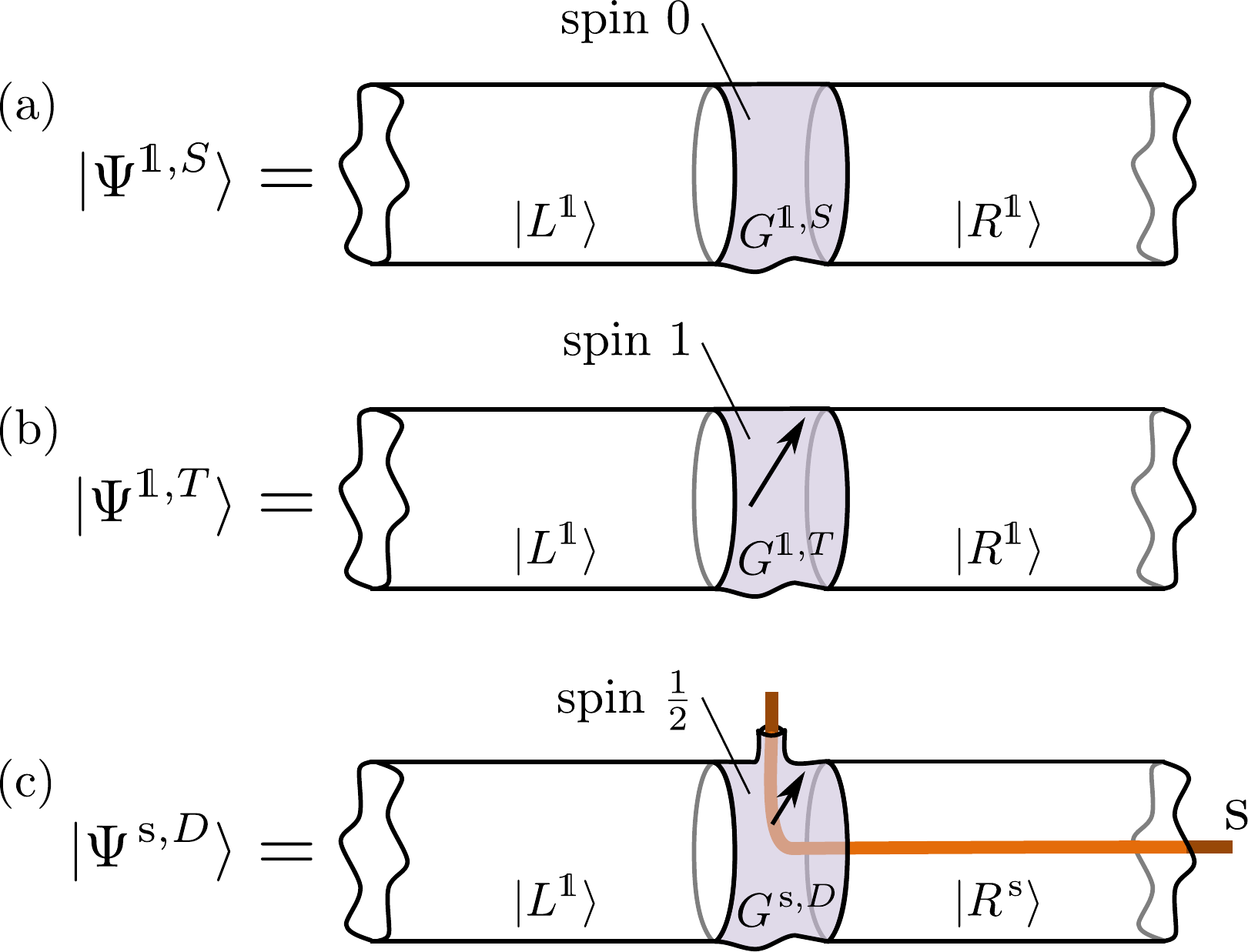}
  \caption{(Color online) An intuitive picture illustrating the ansatz $\ket{\Psi^{a,\mu}}$ for different types of excitations. (a) The Schmidt coefficients $\lambda^\id$ from Fig.~\ref{fig:gs}(a) are replaced by a matrix $G^{\id,S}$ to form a singlet excitation $\ket{\Psi^{\id,S}}$. (b) An ansatz for the triplet excitation $\ket{\Psi^{\id,T}}$ is obtained by inserting matrix $G^{\id,T}$. Matrices $G^{\id,S}$ and $G^{\id,T}$ are energy optimized under different set of constraints discussed in the text. (c)~In turn, an ansatz for the topologically non-trivial excitation $\ket{\Psi^{\sem,D}}$ is constructed by gluing two semi-infinite cylinders, one with identity flux and the other with semion flux inside. The semion flux propagating in the right part of the cylinder creates a spin-1/2 semionic excitation. All the excitations are localized in the shaded region and have a well-defined transverse momentum $k_y$.
  \label{fig:ansatz} }
\end{figure}

Let us now replace the Schmidt coefficients $\lambda^\id$ by a matrix $G^{\id,T}$, which is constrained in a way that $G^{\id,T}_{\alpha\beta} \neq 0$ only if indices $\alpha$ and $\beta$ are such that $\ket{L^\id_\alpha}\ket{R^\id_\beta}$ forms a state with $J=1$ and $J_z=1$. The above constraint leads to an ansatz for a spin-$1$ excitation
\begin{equation} \label{eqn:triplet}
\ket{\Psi^{\id,T}} = \sum_{\alpha,\beta} G^{\id,T}_{\alpha\beta} \ \ket{L^\id_\alpha} \ket{R^\id_\beta} \ .
\end{equation}
In the following, $\ket{\Psi^{\id,T}}$ will be referred to as triplet excitation. Strictly speaking, $\ket{\Psi^{\id,T}}$ is one of the three states of an SU(2) triplet. States with $J_z = 0$ and $J_z = -1$ may be constructed in a similar way. Correspondingly, matrix $G^{\id,T}$ is obtained as a result of minimizing the energy of $\ket{\Psi^{\id,T}}$.

Eqs. (\ref{eqn:singlet},\ref{eqn:triplet}) represent topologically trivial (neutral) excitations above the ground state with identity flux. Having access to the ground state with the semion propagating inside the cylinder, we can generalize the ansatz (\ref{eqn:ansatz}) to describe topologically non-trivial excitations~\cite{cincio2012-1,zaletel2012,zaletel2013}. The ansatz for such an excitation takes the form
\begin{equation} \label{eqn:semion}
\ket{\Psi^{\sem,D}} = \sum_{\alpha,\beta} G^{\sem,D}_{\alpha\beta} \ \ket{L^\id_\alpha} \ket{R^\sem_\beta} \ .
\end{equation}
Here, we force the state $\ket{\Psi^{\sem,D}}$ to have well-defined semion flux inside the right semi-infinite cylinder and identity flux in the left part of the cylinder. Semion flux is seen to create a spin-1/2 topological defect around the position of the matrix $G^{\sem,D}$. In this setting, the matrix $G^{\sem,D}$ is chosen to minimize the energy, subject to a normalization constraint only.

Fig.~\ref{fig:ansatz} provides an intuitive illustration of the ansatz for the different types of excitations described above. Appendix~\ref{app:mps} discusses the specific realization of that ansatz in which the Schmidt vectors $\ket{L_\alpha}$ and $\ket{R_\alpha}$ are given by matrix-product states. In that setting, we explain how to compute the energy and momentum of topologically trivial and non-trivial excitations. Appendix~\ref{app:mps} also describes the computation of overlaps between ground states and excited states on an infinite cylinder.

\section{Results}
\label{sct:results}

\begin{figure}
  \includegraphics[width=0.99\columnwidth]{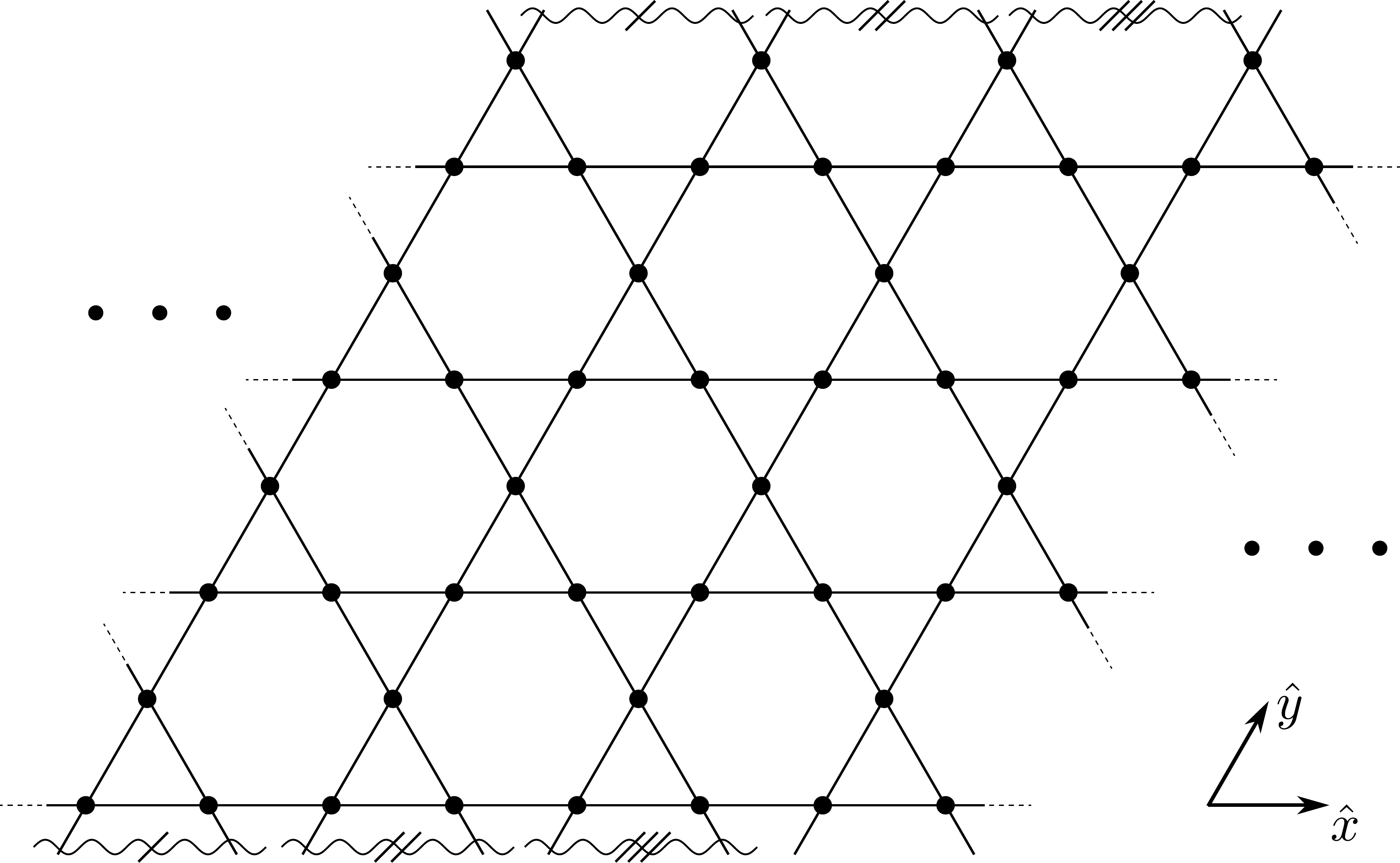}
  \caption{Section of an infinite XC8-4 cylinder. Wavy lines denote boundary conditions in the $\hat{y}$ direction of the cylinder.
  \label{fig:kagome} }
\end{figure}

We have applied the methods discussed in Section~\ref{sct:methods} -- specified in the context of matrix-product states
in Appendix~\ref{app:mps} -- to the model given in Eqn.~\eqnref{eqn:H}. Throughout this work, we use a cylinder in the
XC8-4 geometry, as shown in Fig.~\ref{fig:kagome} and quote results obtained for a bond dimension of up to \mbox{$\chi=4096$}.

\begin{figure}
  \includegraphics{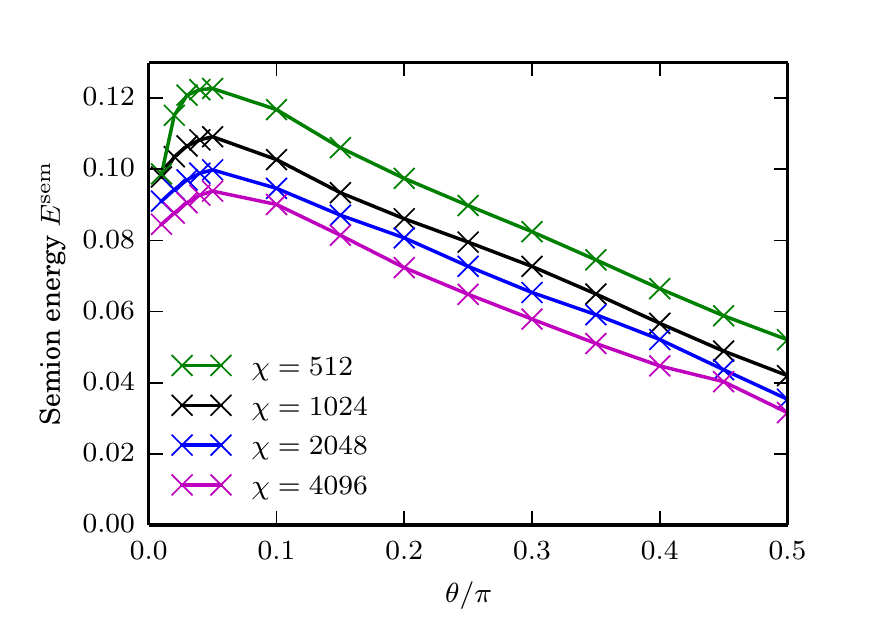}
  \caption{(Color online) Energy of a single, isolated semionic excitation above the ground state of \eqnref{eqn:H}.
  Results are obtained for a cylinder of width 8 with a bond dimension of $\chi = 4096$. \label{fig:semionenergy} }
\end{figure}

In Fig.~\ref{fig:semionenergy}, we show the energy of a single semion, $E^\mathrm{sem}$, throughout
the phase diagram of Hamiltonian~\eqnref{eqn:H}.
We observe that the semion energy increases as the model is tuned from the fully chiral point at $\theta=\pi/2$
to the Heisenberg point $\theta=0$. For all values of $\theta$, the semion energy decreases slowly as the bond
dimension is increased (in the figure, note that the bond dimensions are spaced exponentially).
Crucially, the semion energy seems to remain finite throughout the entire phase
diagram, including at very small values of $\theta$ in the range $0 < \theta \leq 0.05\pi$ where we expect
a phase transition.

\begin{figure}
  \includegraphics{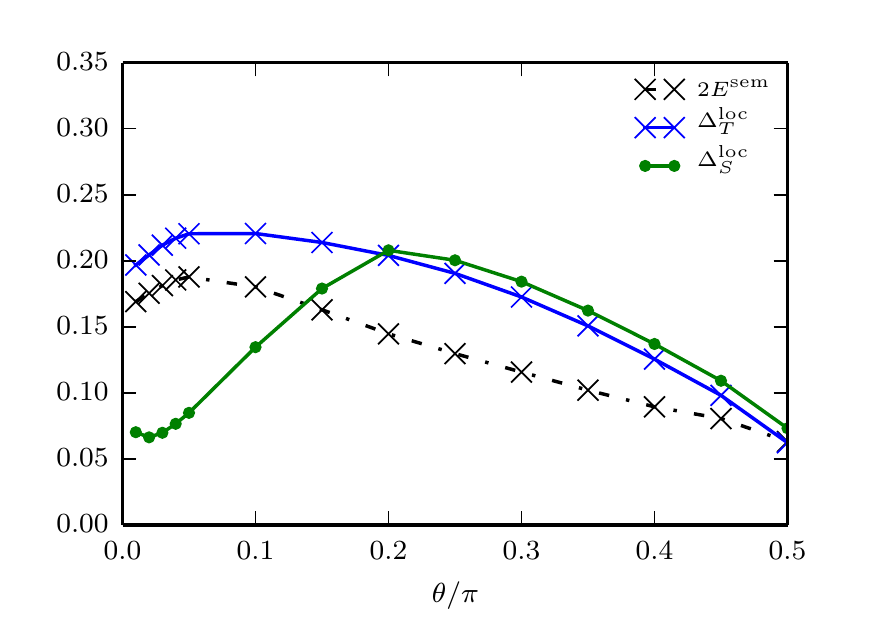}
  \caption{(Color online) Energies of low-lying topological and trivial excitations on a XC8-4 cylinder with
  $\chi=4096$. Excitation energies shown here are measured in the identity sector; results for the semion sector
  are very similar.\label{fig:gaps} }
\end{figure}

In Fig.~\ref{fig:gaps}, we show, in addition to the energy cost of two semions at infinite distance, $2 E^\mathrm{sem}$,
the variational bounds on the energies of local, topologically trivial excitations of spin $J=0$ (singlet) and $J=1$ (triplet),
which we will below denote as $\Delta^\mathrm{loc}_S$ and $\Delta^\mathrm{loc}_T$, respectively. Each of these quantities can be extracted
for excitations above the ground state in the identity and semion sectors; our results show that these gaps agree within
reasonable numerical accuracy and we will thus focus on the gaps above the identity sector in the following.

A key observation is that at the fully chiral point
$\theta = \pi/2$, where the first term of Eqn.~\eqnref{eqn:H} -- the Heisenberg coupling -- vanishes, the energies of a local
singlet and triplet excitation approximately coincide with twice the energy of a semion, that is
\begin{equation}
\Delta^\mathrm{loc}_S \approx \Delta^\mathrm{loc}_T \approx 2 E^\mathrm{sem}.
\end{equation}
This suggests the interpretation that at least for $\theta = \pi/2$, the semions interact very weakly at short distances
and the local singlet and triplet excitations, which are topologically trivial, consist of a pair of semions. This is
consistent with the two possible SU(2) fusion outcomes for a semion of spin $J=1/2$, given by the usual SU(2) spin
addition rule $1/2 \otimes 1/2 = 0 \oplus 1$. In the
following, we will examine this assertion in more detail and in particular carefully examine whether it extends away
from $\theta = \pi/2$.

In the data shown in Fig.~\ref{fig:gaps}, we see that as the Hamiltonian is tuned away from the purely chiral point,
the energies of local excitations behave markedly different.
Focusing first on the behavior of the energies for topologically trivial excitations $\Delta^\mathrm{loc}_S$ and $\Delta^\mathrm{loc}_T$, we find
that they remain finite throughout the phase diagram. This is to be expected even in the case of a continuous
phase transition from the chiral spin liquid to a topologically distinct gapped phase around the Heisenberg point
as the transition is rounded to a crossover due to the finite size of the system, and the extent of the excitation
being considered is limited due to the finite bond dimension of the MPS.
We also note that the upper bounds on the excitation gaps given by $\Delta^\mathrm{loc}_S$ and $\Delta^\mathrm{loc}_T$
shown in Fig.~\ref{fig:gaps} do not seem to be tight, as we can construct a
(non-local) excitation of two well-separated semions, which, assuming that interactions between anyons
decay with distance, will have energy $2 E^\mathrm{sem}$.

However, the data also clearly shows that while the triplet gap $\Delta^\mathrm{loc}_T$ remains quantitatively unchanged across
the transition, the singlet gap $\Delta^\mathrm{loc}_S$ drops significantly as the transition is approached. This would be consistent
with a closing of the singlet gap in the thermodynamic limit, while the triplet gap remains finite across the transition
even in the thermodynamic limit. Unfortunately, reliable extrapolation to the thermodynamic limit in critical region
is beyond the capabilities of current DMRG methods.

\begin{figure}
  \includegraphics{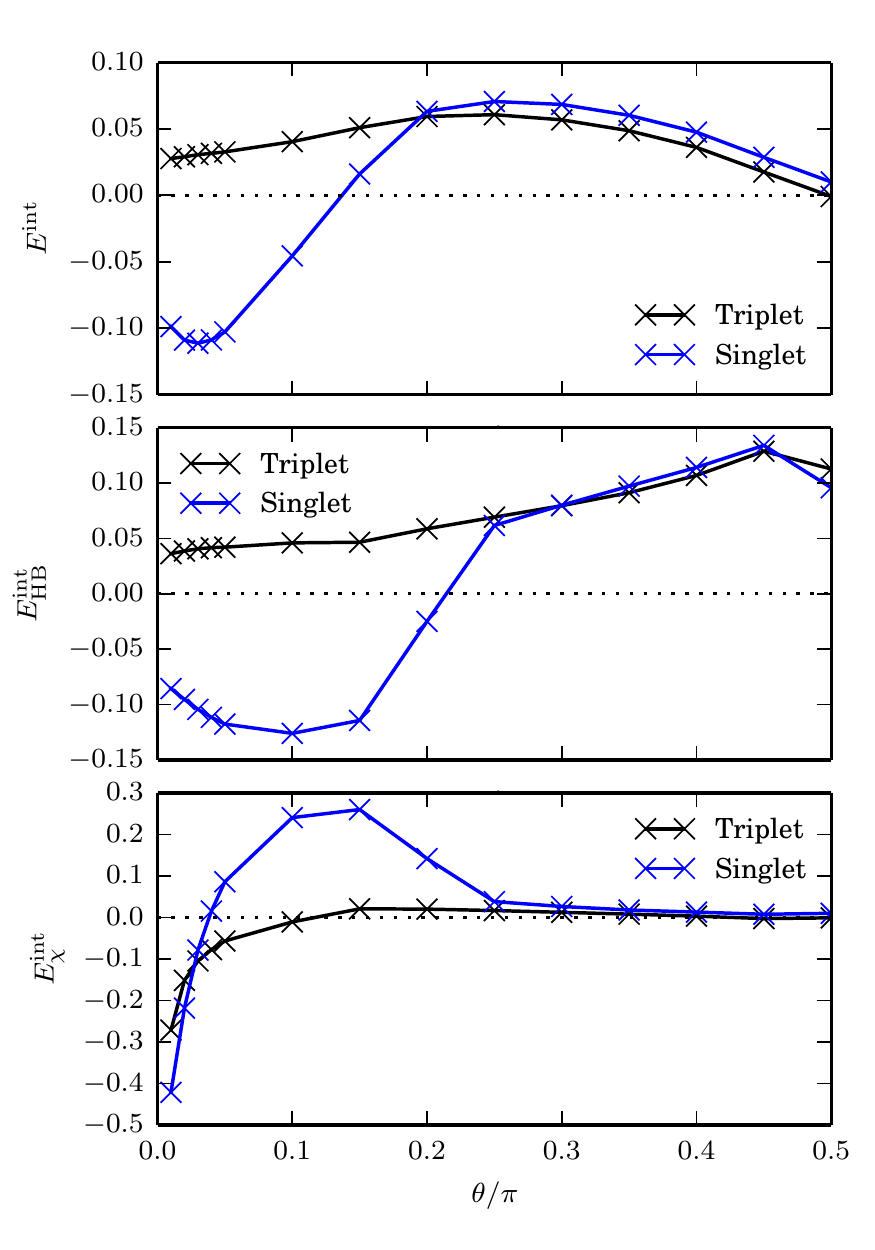}
  \caption{(Color online) Interaction energies extracted using the ansatz~\eqnref{eqn:int}. Results are
  obtained using $\chi=4096$. \label{fig:interactions} }
\end{figure}

The interpretation of the local singlet and triplet excitation as a pair of semions naturally suggests an
interpretation of the difference between the energy of the trivial excitation and that of two semions
as an interaction energy between semions,
\begin{equation} \label{eqn:int}
\begin{split}
E^\mathrm{int}_S \equiv & \ \Delta^\mathrm{loc}_S - 2 E^\mathrm{sem} \ , \\
E^\mathrm{int}_T \equiv & \  \Delta^\mathrm{loc}_T - 2 E^\mathrm{sem} \ .
\end{split}
\end{equation}
This can be thought of as the energy cost of bringing two infinitely distant semions close to each other (on the scale
of the correlation length), or conversely as the energy gain from locally creating a pair of semions and bringing them
to a distance from each other that is much larger than the correlation length. When $E^\mathrm{int}$ is positive, the
two semions thus interact repulsively, while they interact attractively for $E^\mathrm{int} < 0$.

Our numerical results for this quantity are shown in the top panel of Fig.~\ref{fig:interactions}. At $\theta=\pi/2$,
we have that $E^\mathrm{int} \approx 0$ both in the singlet and the triplet channel. For $\pi/4 < \theta \leq \pi/2$, we find
that the interaction energy becomes repulsive, $E^\mathrm{int} > 0$, in both channels.
As the transition regime is approached further, we observe drastically different behavior in the singlet and
triplet channels: while the interaction energy in the triplet channel remains qualitatively unchanged as the transition
is approached, the interaction energy in the singlet channel drops and eventually changes its sign at $\theta \approx 0.15\pi$.
Note that in Fig.~\ref{fig:gaps} this behavior is reflected in the fact that $\Delta^\mathrm{loc}_S$ crosses $2 E^\mathrm{sem}$ around
that value of $\theta$.

\begin{figure}
  \includegraphics{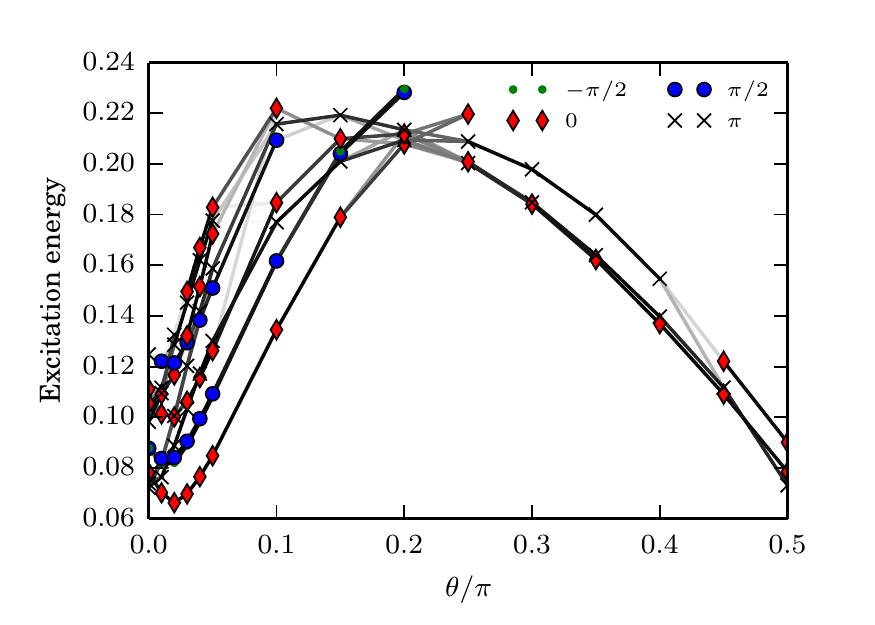}
  \includegraphics{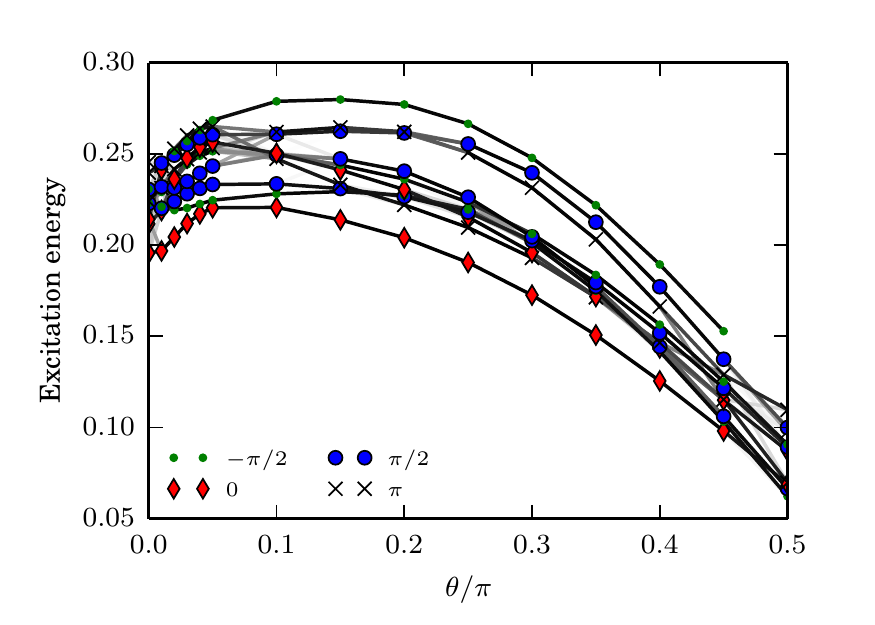}
  \caption{(Color online) Singlet (top panel) and triplet (bottom panel) excitation energies for $\chi=4096$. Here, the
  symbols denote the $k_y$ momentum of the excitation, and the thickness of the lines indicates the overlap
  between excited states. \label{fig:fid} }
\end{figure}

The sudden crossover in $E^\mathrm{int}_S$ suggests a further investigation of the question whether the interpretation
of the localized singlet excitation as a pair of semions remains valid around $\theta \approx \pi/4$. To this end,
we have calculated the fidelities of low-lying singlet and triplet excitations, as shown in Fig.~\ref{fig:fid}.
In the figure, data points indicate the energy of low-lying local excitations, and the thickness and shading
of the lines connecting them indicate the fidelity of a pair of excitations.
In the triplet sector, we find that the lowest excitation, which carries transverse momentum $k_y=0$ forms
a branch of excitations with high overlap that is well-separated from the rest of the spectrum, both in terms of
its energy as well as its fidelity. This strongly suggests that the interpretation of the triplet as a pair of
semions holds not only at the fully chiral point $\theta=\pi/2$, but rather holds throughout the entire phase
diagram up to the transition.
In contrast to this, the spectrum of excitations in the singlet sector shows a major reorganization around
$\theta=\pi/4$, coincidental with the regime where the excitation energy changes drastically. In particular,
the branch of the lowest excitation for $\theta > \pi/4$ moves up in the spectrum for $\theta < \pi/4$. This
must be taken as indication that the interpretation of the local singlet excitation as a pair of semions must
become unreliable in the vicinity of $\theta = \pi/4$.

We can further resolve this data by calculating the variational energies of the ansatz states independently
for the Heisenberg and chiral interaction, which we will now denote as $E^\mathrm{int}_\mathrm{HB}$ and
$E^\mathrm{int}_\chi$, respectively. These quantities are shown in the middle and bottom panels of Fig.~\ref{fig:interactions}.
The results show that the interaction energy due to the chiral term for the triplet excitation is very small
throughout the phase diagram; in the singlet channel, the interaction energy increases sharply around $\theta = \pi/4$.
For the interaction due to the Heisenberg term, we observe that it remains qualitatively the same in the triplet case
throughout the entire phase diagram, while in the singlet case it shows a sharp decrease around $\theta = \pi/4$,
where the Heisenberg term becomes the dominant term in the Hamiltonian. This is also consistent with
the major reorganization of the excitation spectrum at $\theta = \pi/4$ driven by the presence of a Heisenberg
of comparable strength to the chiral term.

\section{Conclusion}

We have described matrix-product state techniques to study the excitations of a topological phase on infinite
cylinders. These techniques allow us to obtain both topologically trivial excitations
as well as non-trivial excitations, which are obtained as domain walls between cylinders of different
topological flux. In our approach, the local excitations are confined to within roughly one correlation
length of the MPS, and energies obtained in this way thus provide an upper bound on the energy of
the excitations. These bounds could be improved by allowing larger spatial extent of the excitations.

We apply these techniques to a Heisenberg model where a chiral three-spin term breaks time-reversal symmetry
explicitly and favors a chiral spin liquid phase. The model we study interpolates between the chiral spin liquid
in the limit of sufficiently strong time-reversal symmetry breaking, and the time-reversal symmetric phase
that surrounds the unperturbed Heisenberg point. We show that in the limit where only the three-spin chiral
term is present, the low-lying singlet and triplet excitations are given as pairs of weakly interacting semions.
As the system is driven towards the Heisenberg point, the semions begin to interact and the energetics of the
low-lying excitations change significantly. We observe that for the finite systems available to our numerical
methods, both the energy of a single semion as well as the energy of a local triplet excitation remain finite
across the entire phase diagram. The energy of a local singlet excitation, on the other hand, drops sharply
as the transition is approached, which is accompanied by a major reorganization of the low-lying singlet
excitations in the chiral phase.

\acknowledgements

We acknowledge useful discussions with M. Barkeshli and M. Zaletel. L.C. and G.V. acknowledge support by the John Templeton Foundation. G.V. also acknowledges support by the Simons Foundation (Many Electron Collaboration). This research was supported in part by Perimeter Institute for Theoretical Physics. Research at Perimeter Institute is supported by the Government of Canada through Industry Canada and by the Province of Ontario through the Ministry of Research and Innovation.

\appendix

\section{Matrix-product ansatz for localized excited states}
\label{app:mps}
In this Appendix we discuss a matrix-product state ansatz for localized excited states outlined in Section~\ref{sct:methods}.

\begin{figure}[!t]
  \includegraphics[width=0.99\columnwidth]{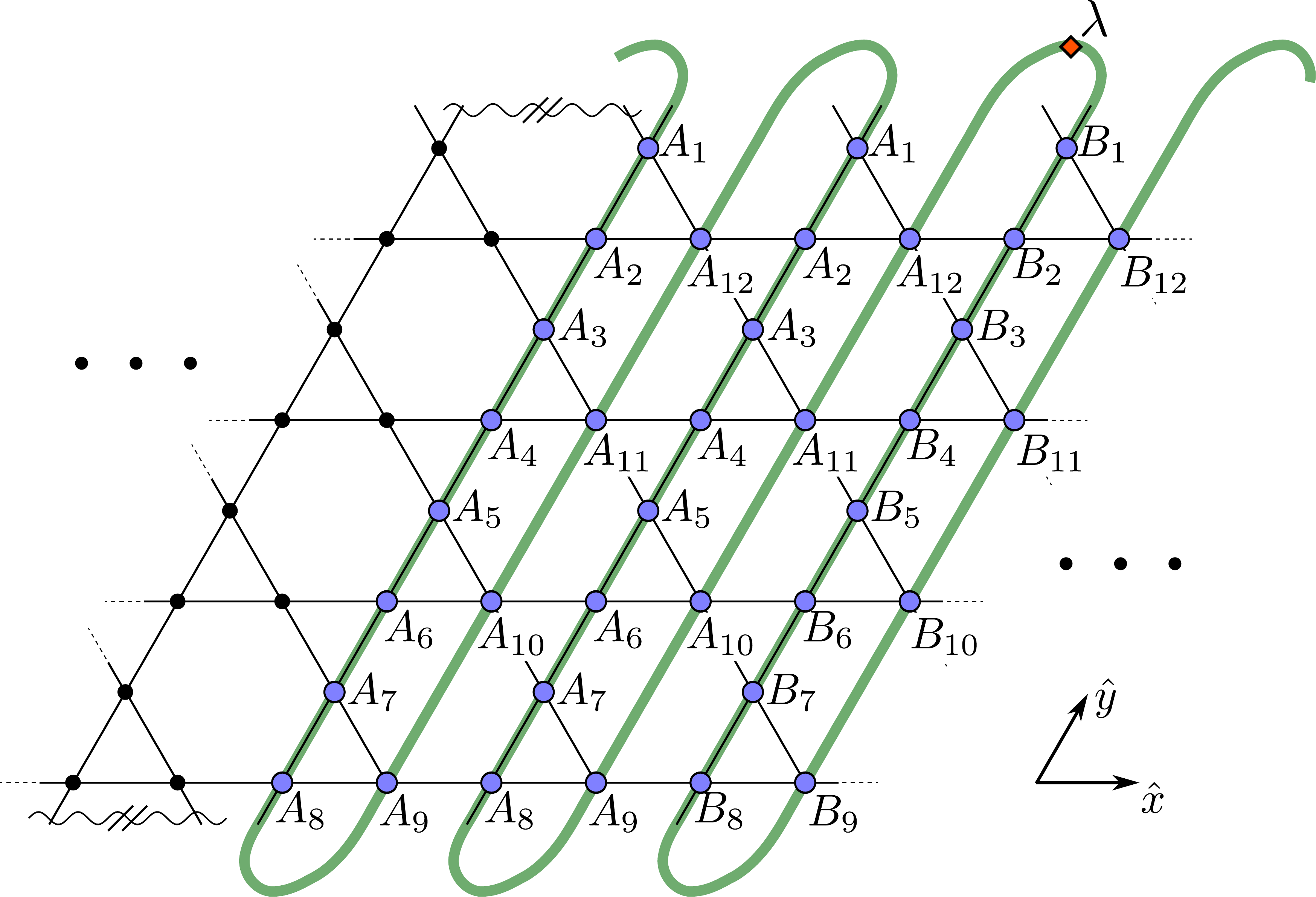}
  \caption{(Color online) Matrix-product state representation of a state on an infinite XC8-4 cylinder with finite width $W=4$ lattice unit cells. Green line represents the mapping of an infinite cylinder to a 1d path. The MPS ansatz is composed of a unit cell of $n=12$ tensors. Boudary conditions are marked with a wavy line $\begin{matrix}\protect\includegraphics[height=0.17cm]{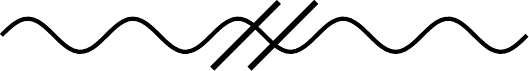}\end{matrix}$. \label{fig:mps} }
\end{figure}

The system under consideration consists of a cylinder with infinite length and finite width $W$, as measured by the lattice unit cells. Fig.~\ref{fig:kagome} in Section~\ref{sct:methods} shows the Kagome lattice with our choice of boundary conditions in the transverse direction for $W=4$. In this setup, the Hamiltonian remains translationally invariant in both longitudinal and transverse directions. However, a matrix-product ansatz for the state on an infinite cylinder is translationally invariant in $\hat{x}$ direction only. Translational invariance in the $\hat{y}$ direction will only be approximately recovered through energy minimization. The MPS ansatz considered here for ground states $\ket{\Phi}$ and excited states $\ket{\Psi}$ consists of a unit cell of $n$ different tensors that is repeated throughout an infinite cylinder. We denote the MPS tensors as $\{ A_i \}$ and $\{  B_i  \}$, where $i=1,\ldots,n$ enumerates the tensors on inequivalent sites within the unit cell, and denote the bond dimension of each tensor as $\chi$. Tensors $\{ A_i \}$ ($\{  B_i  \}$) correspond to the left (right) semi-infinite part of a cylinder, see Fig.~\ref{fig:mps}.

The Schmidt decomposition of the ground state $\ket{\Phi}$ on an infinite cylinder takes the form
\begin{equation} \label{eqn:Schmidt_gs}
\ket{\Phi} = \sum_\alpha \lambda_\alpha \ket{L_\alpha} \ket{R_\alpha} \ ,
\end{equation}
where the Schmidt vectors $\ket{L_\alpha}$ and $\ket{R_\alpha}$ are represented by the MPS tensors (compare with Fig.~\ref{fig:mps}):
\begin{eqnarray}
\ket{L_\alpha} \ &=& \ \ingr{1.0}{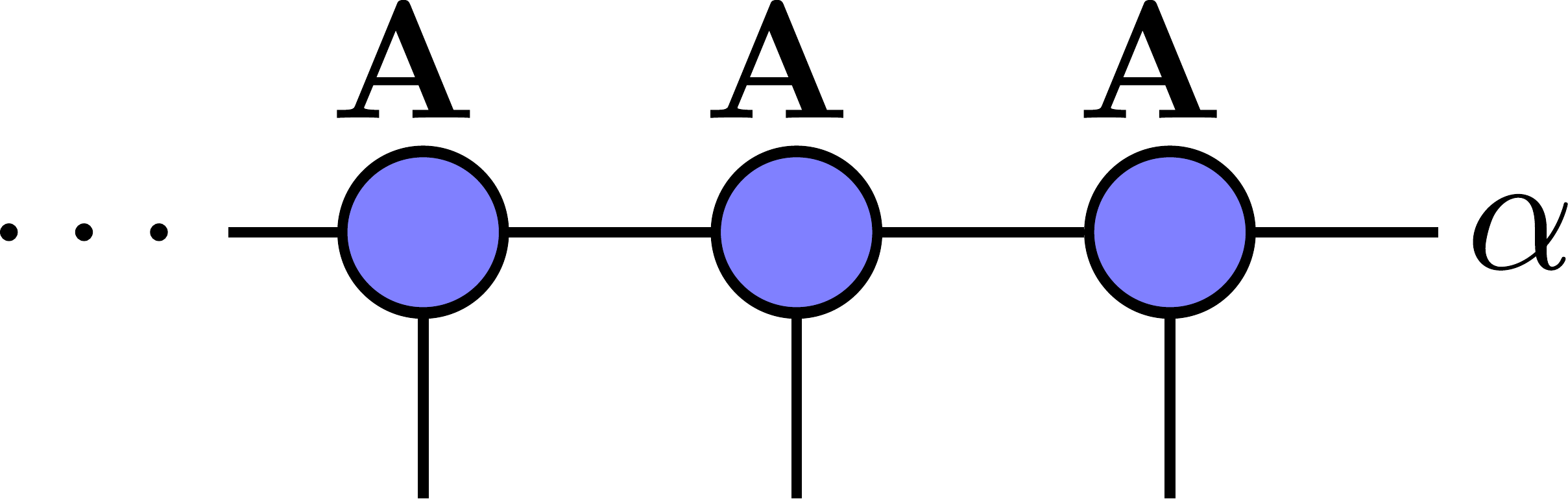} \label{eqn:SchmidtL} \ ,\\
\ket{R_\alpha} \ &=& \ \ingr{1.0}{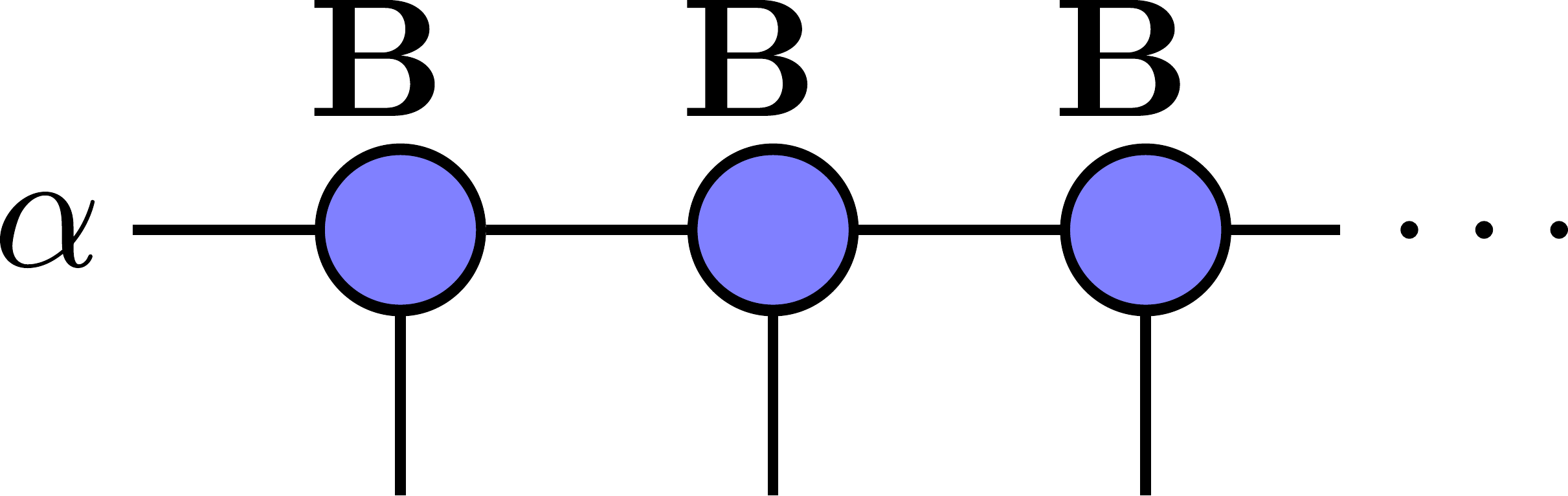} \label{eqn:SchmidtR} \ .
\end{eqnarray}
We emphasize that, in (\ref{eqn:SchmidtL}), $\bf A$ represents not just one single tensor but the entire unit cell of tensors $A_i$:
\begin{equation}
\ingr{1.0}{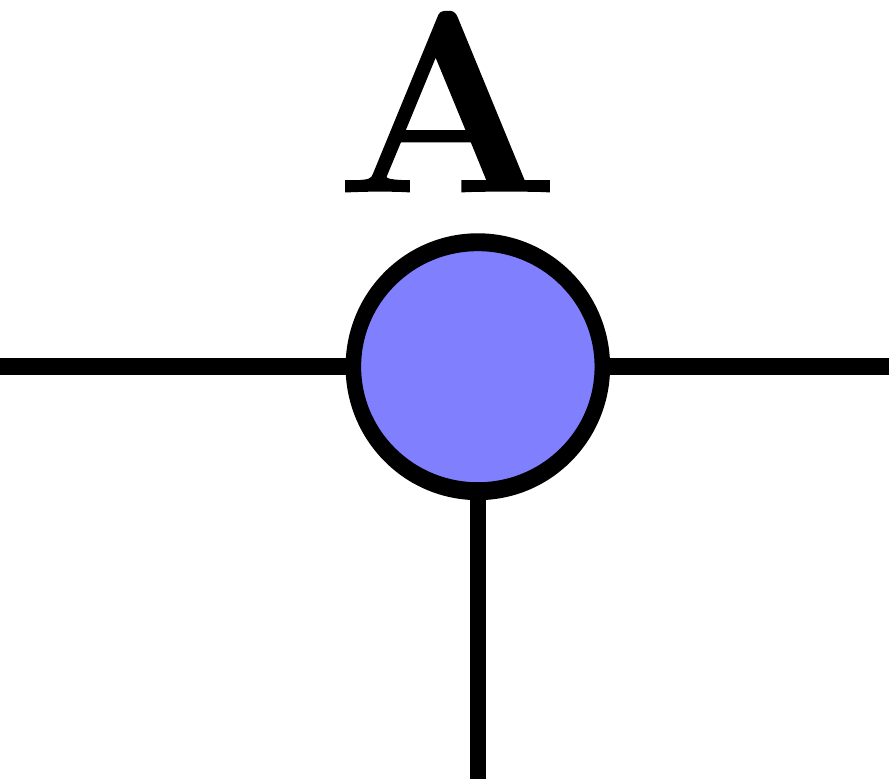} \ = \ \ingr{1.0}{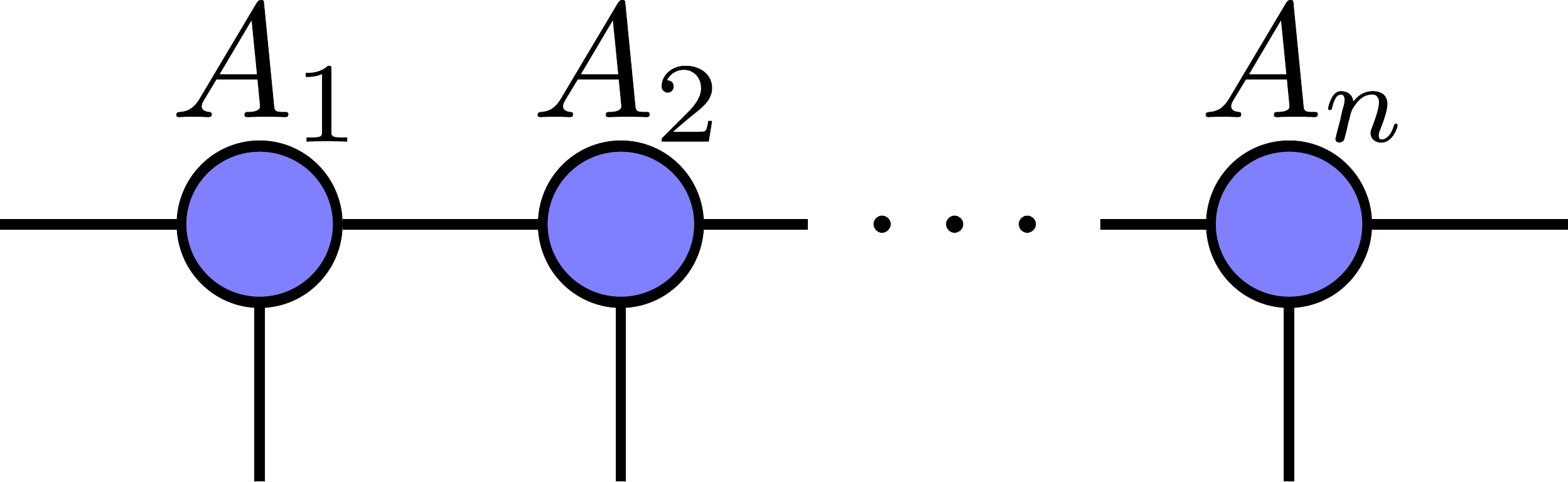}
\end{equation}
and similarly for tensor $\bf B$ in (\ref{eqn:SchmidtR}). In the canonical MPS representation of $\ket{L_\alpha}$ and $\ket{R_\alpha}$, tensors $\bf A$ and $\bf B$ are related by ${\bf B} = \lambda^{-1} {\bf A} \lambda$, where $\lambda$ is the diagonal matrix of the Schmidt coefficients in (\ref{eqn:Schmidt_gs}).

As described in Section~\ref{sct:methods}, an excited state is constructed by replacing the Schmidt coefficients $\lambda$ in (\ref{eqn:Schmidt_gs}) by a matrix $G$:
\begin{equation} \label{eqn:mpsExc}
\ket{\Psi} \ = \  \ingr{1.0}{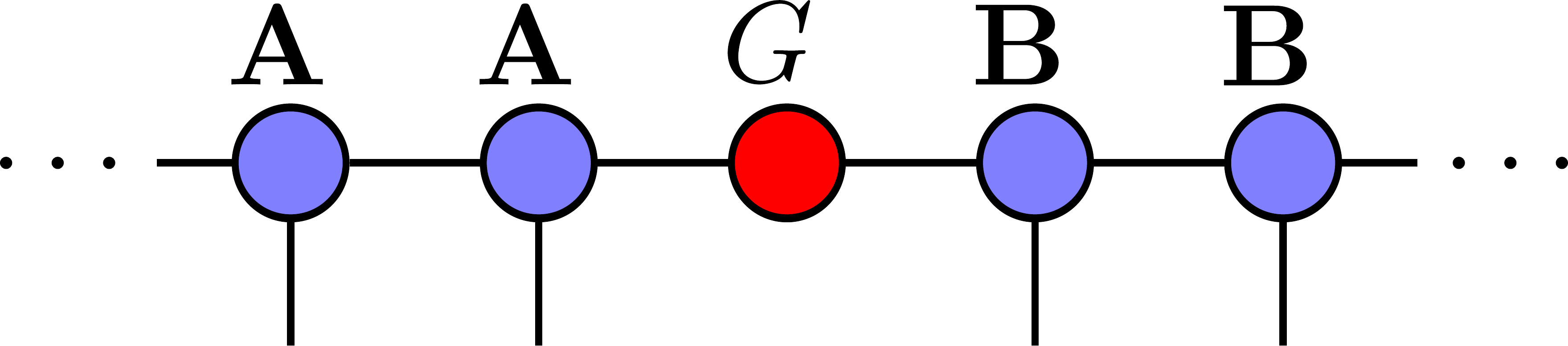} \ ,
\end{equation}
where $G$ is a $\chi \times \chi$ bond matrix without physical index. Normalization of $\ket{\Psi}$ requires that $\braket{\Psi}{\Psi} \equiv \Tr G^\dagger G = 1$. The matrix $G$ is constrained by $\braket{\Phi}{\Psi} \equiv \sum_\alpha \lambda_\alpha G_{\alpha\alpha} = 0$ to assure that the excited state $\ket{\Psi}$ is orthogonal to the ground state $\ket{\Phi}$. Forcing additional constraints on the matrix $G$ discussed below, leads to an ansatz for a singlet $\ket{\Psi^{\id,S}}$ or a triplet $\ket{\Psi^{\id,T}}$ excitation with corresponding matrices $G^{\id,S}$ and $G^{\id,T}$.

Additionally, a topologically non-trivial excitation is obtained by connecting two semi-infinite cylinders carrying identity and semion flux. Following Eq.~(\ref{eqn:semion}), the MPS ansatz is given by
\begin{equation}
\ket{\Psi^{\sem,D}} \ = \ \ingr{1.045458}{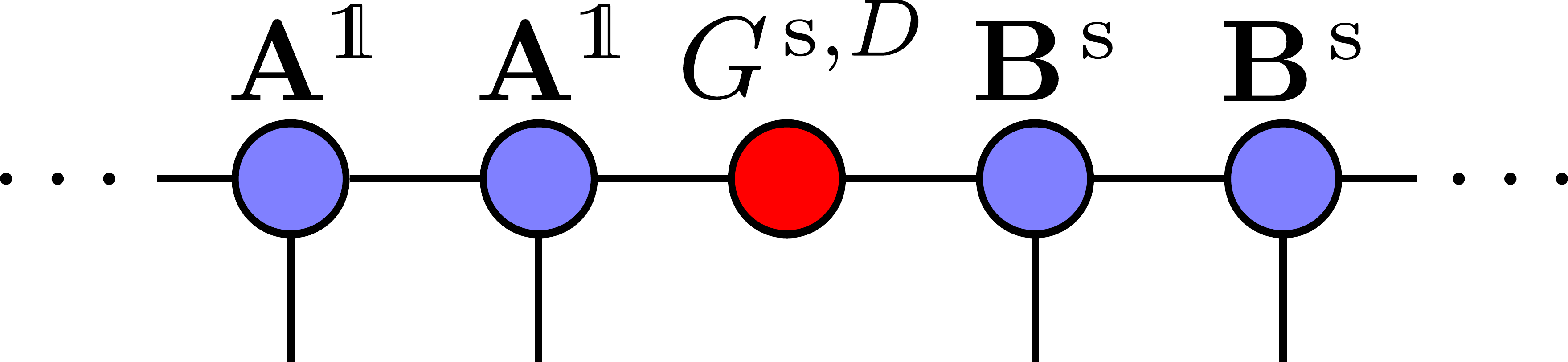} \ ,
\end{equation}
where ${\bf A}^\id$ forms the Schmidt vectors $\ket{L^\id_\alpha}$ of the ground state $\ket{\Phi_\id}$ having identity flux threading through the cylinder. Similarly, ${\bf B}^\sem$ corresponds to the Schmidt vectors $\ket{R^\sem_\alpha}$ of $\ket{\Phi_\sem}$ carrying the semion flux.

In order to build the above localized excitations with SU(2) charge $J$ and $J_z = J$ it is enough to restrict our attention to the U(1) subgroup generated by $J_z$. Correspondingly, we have used an MPS that is made of U(1) symmetric tensors. In that case, an index $\alpha$ labeling Schmidt vectors $\ket{L_\alpha}$ and $\ket{R_\alpha}$ in (\ref{eqn:SchmidtL},\ref{eqn:SchmidtR}) is represented by a pair $\alpha = (c,i)$, where $c$ is a U(1) charge and $i$ labels different Schmidt vectors carrying the same charge $c$. The matrix $G$ in (\ref{eqn:mpsExc}) is U(1) symmetric, which means that it maps charge $c$ to (possibly different) charge $c + \Delta c$. Technically, this implies that $G$ has a block structure in which the only non-zero elements $G_{(c,i),(c',i')}$ are such that $c' = c - \Delta c$, where $\Delta c$ is the same for all possible values of $c$. In this setting, the excited state (\ref{eqn:mpsExc}) is seen to carry U(1) charge $\Delta c$.

A singlet excitation $\ket{\Psi^{\id,S}}$ is obtained by choosing a block structure with $\Delta c = 0$ and ensuring that the resulting excited state does not belong to a larger multiplet. This requirement can be verified numerically by computing the expectation value of $(\pvec{J})^2 = J^2_x + J^2_y + J^2_z$, where $\vec{J}$ is the total spin operator. In practice, we only need to evaluate $(\pvec{J})^2$ on a finite, cylindrical region surrounding the matrix $G^{\id,S}$, denoted here by $\mathcal{A}$. The result is then compared against $(\vec{J}_\mathcal{A})^2$ computed for the ground state which is known to be a singlet. The difference $\bra{\Psi} (\vec{J}_\mathcal{A})^2 \ket{\Psi} - \bra{\Phi} (\vec{J}_\mathcal{A})^2 \ket{\Phi}$ gives $J(J+1)$, where $J$ is the total spin of an excitation. The reason to consider the difference instead of $\bra{\Psi} (\vec{J}_\mathcal{A})^2 \ket{\Psi}$ is that on a finite region $\mathcal{A}$, $\bra{\Psi} (\vec{J}_\mathcal{A})^2 \ket{\Psi}$ contains contributions coming from the boundary of $\mathcal{A}$. However, it follows from the construction of the excited state (\ref{eqn:mpsExc}), that the same boundary contribution is present in $\bra{\Phi} (\vec{J}_\mathcal{A})^2 \ket{\Phi}$.

In order to access a $J_z = 1$ state of a triplet excitation $\ket{\Psi^{\id,T}}$, matrix $G^{\id,T}$ is constrained with $\Delta c = 1$; then one must verify that such an excitation does not belong to some larger multiplet, which can be achieved in a fashion analogous to the above prescription for the singlet excitations.

In the construction of a topologically non-trivial excitation $\ket{\Psi^{\sem,D}}$, matrix $G^{\sem,D}$ has a peculiar block form. In this case, the indices $\alpha$ and $\beta$ of the Schmidt vectors $\ket{L^\id_\alpha}$ and $\ket{R^\sem_\beta}$ carry SU(2) quantum numbers that fuse to half-integer representations of SU(2). In the identity topological sector, $\ket{L^\id_\alpha}$ carries integer $J$, while in the semion sector, $\ket{R^\sem_\beta}$ is associated with half-integer $J$. Thus, the matrix $G^{\sem,D}$ maps U(1) charge $c$ to U(1) charge $c+\Delta c$, where $\Delta c$ is half-integer. We note that this immediately implies that the semion carries fractional U(1) charge (or equivalently, half-integer $J$).

To obtain the energy of a local excitation, we measure all Hamiltonian terms in a region $\mathcal{A}$ that contains the excitation and is sufficiently large compared to the correlation length, see Fig.~\ref{fig:ansatz}(a)--(c). Performing this for the same region both in the ground state $\ket{\Phi}$ as well as the excited state $\ket{\Psi}$, the energy of the local excitation is given as the energy difference. Assuming that the ground state energy was obtained accurately, the energy difference introduced above provides an upper bound for the corresponding excitation gap. This assumption is reasonable, since DMRG is known to accurately capture the properties of the ground state (which may not be true for the excited state). In the case of a topological defect, the additional subtlety arises that the two ground states $\ket{\Phi_\id}$ and $\ket{\Phi_\sem}$ in general have slightly different energy density (due to a splitting that vanishes exponentially in the circumference of the cylinder $W$). In this case, the energy is compared against the average of the energies of the two ground states in region $\mathcal{A}$. Note also that in this approach, we can evaluate the energy separately for different terms of the Hamiltonian (\ref{eqn:H}). Section~\ref{sct:results} discusses independent contributions due to Heisenberg and scalar spin chirality terms.

In addition, we can obtain the transverse momentum $k_y$, i.e. the momentum related to translations around the cylinder, of a local excitation. This computation is crucially based on the observation that translation $T_y$ by one lattice unit cell in $\hat{y}$ direction is a product of operators $T_0$ acting within one MPS unit cell. As can be seen from Fig.~\ref{fig:mps}, the overlap between MPS tensors $\bf A$, ${\bf A}^\ast$ and the translation $T_0$ is given by
\begin{equation}
\ingr{1.782876}{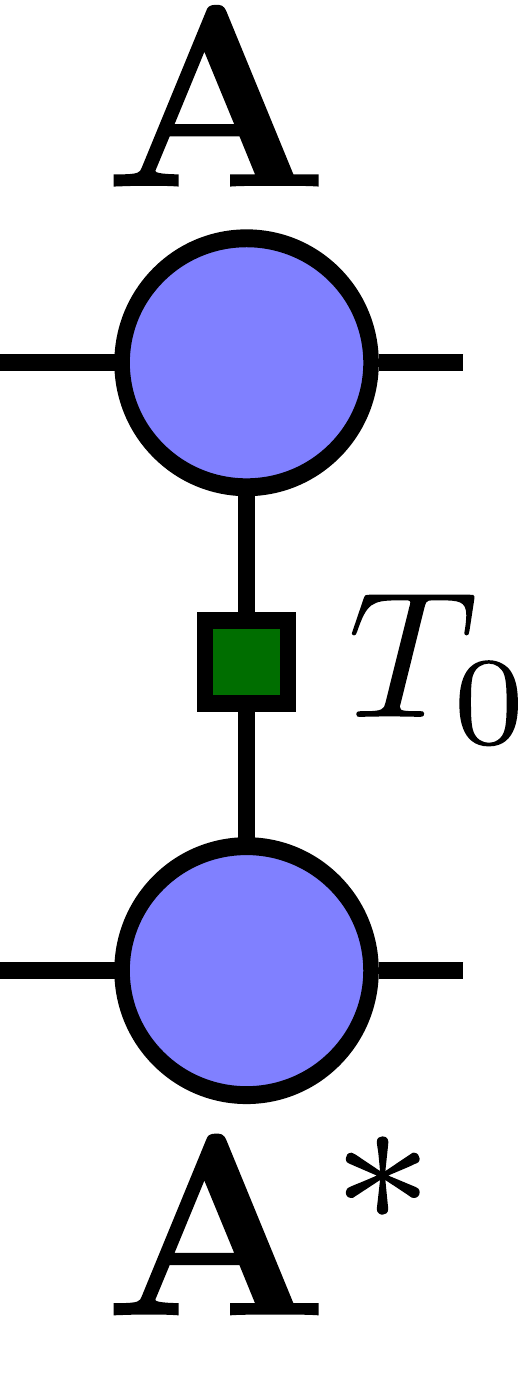} \ = \ \ingr{1.782876}{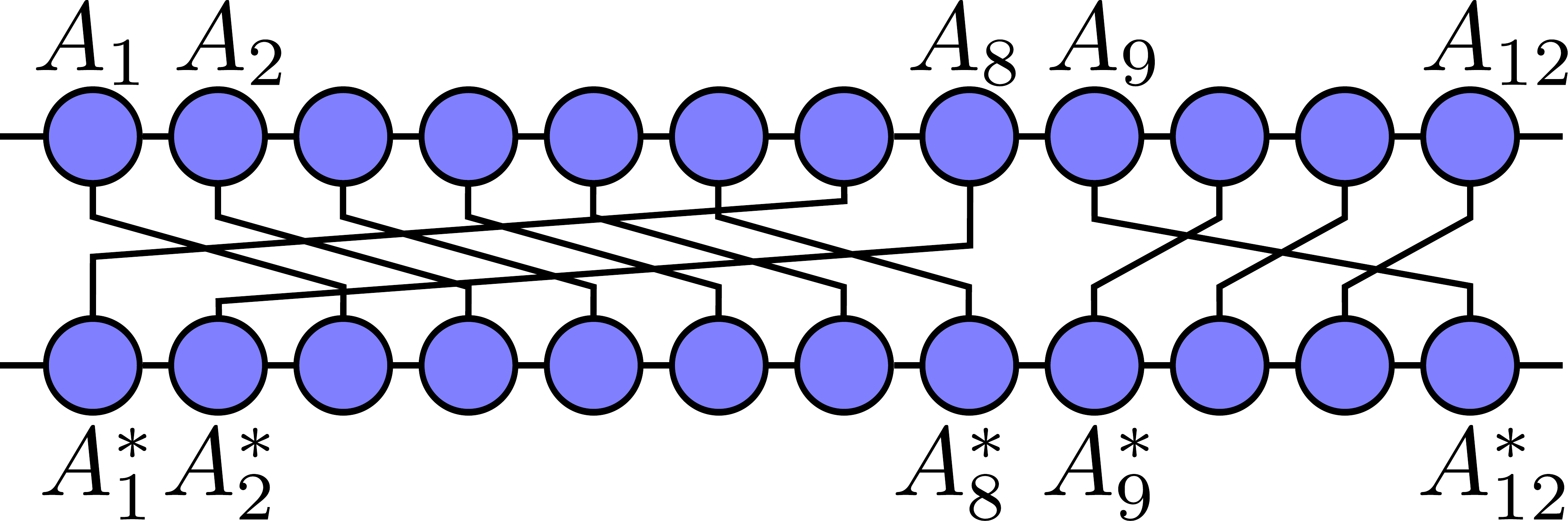} \ .
\end{equation}
This observation allows us to measure the expectation value of translation $T_y$, which for a momentum eigenstate should equal $e^{i k_y}$. The transverse momentum is thus given by
\begin{equation} \label{eqn:mom}
e^{ik_y} \ = \ \ingr{1.719489}{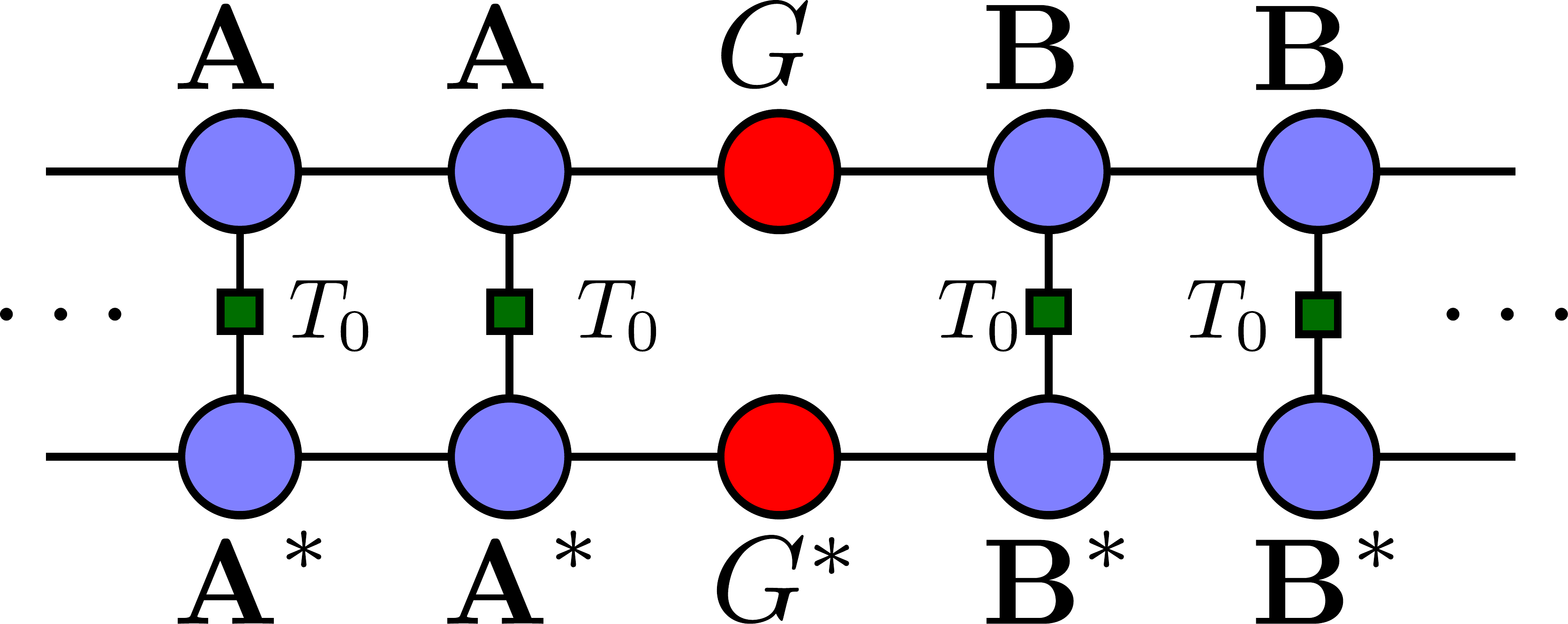} \ = \ \ingr{1.719489}{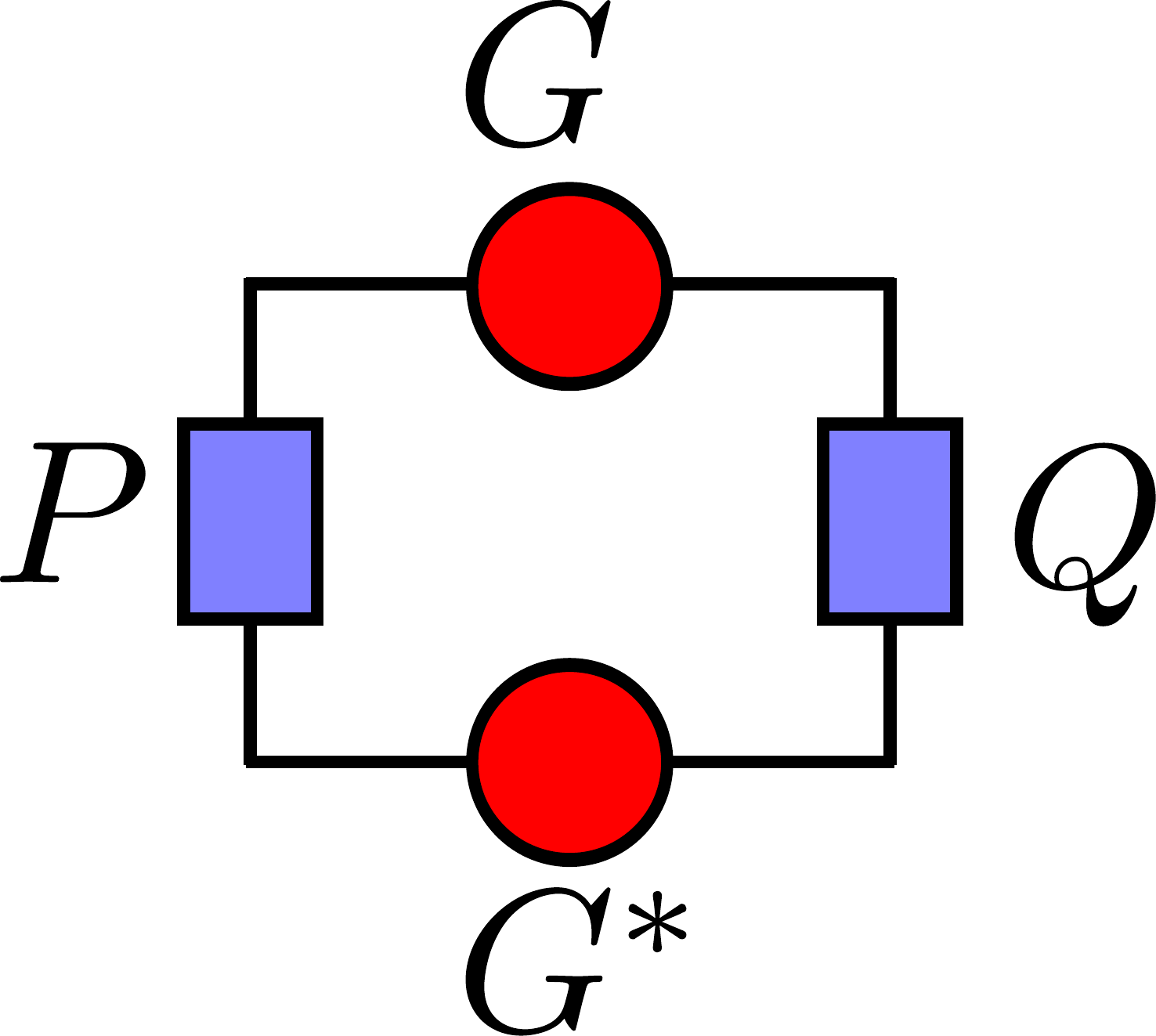} \ .
\end{equation}
As the boundary objects (denoted as $P$ and $Q$ in \eqnref{eqn:mom}) are only defined up to a phase, this expectation value is also only defined up to a phase. However, since the boundary objects are the same for the ground state $\ket{\Phi}$ and excited states $\ket{\Psi}$ that differs only by a bond matrix $G$, we can obtain the momentum of an excited state relative to the ground state, which is assumed to have momentum $k_y = 0$. For a well-converged excited state, the momentum will be close to one of a set of discrete eigenvalues $k_y = \frac{2 \pi}{W} m$, $m=0,\ldots,W-1$. If the bond dimension $\chi$ is insufficiently large, we often find that $k_y$ is not near one of these values, indicating that the state $\ket{\Psi}$ is not converged to an eigenstate of the translation operator $T_y$.

Finally, let us consider two infinite MPSs
\begin{eqnarray}
\ket{\Theta(G)}  & \ = \ & \ingr{1.0}{mps1.pdf} \ ,\\
\ket{\Theta'(G')} & \ = \ & \ingr{1.0}{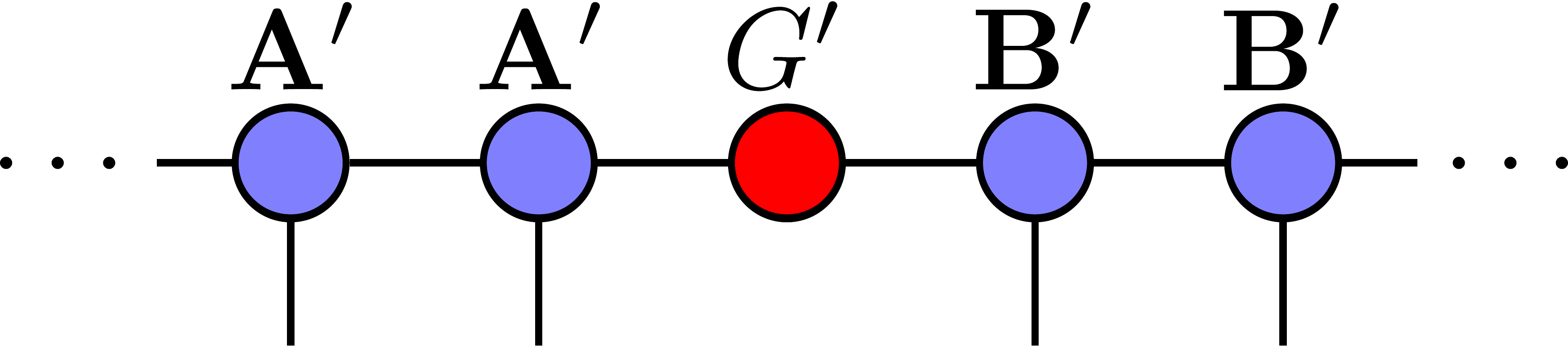} \ .
\end{eqnarray}
States $\ket{\Theta(G)}$ and $\ket{\Theta'(G')}$ may represent localized excitations or ground states. In the latter case, $G_{\alpha\beta} = \delta_{\alpha\beta}\lambda_\alpha$ and $G'_{\alpha\beta} = \delta_{\alpha\beta}\lambda'_\alpha$, where $\lambda_\alpha$, $\lambda'_\alpha$ are the Schmidt coefficients (\ref{eqn:Schmidt_gs}).

The {\it overlap per unit cell} between ground states $\ket{\Theta(\lambda)}$ and $\ket{\Theta'(\lambda')}$ is defined as the absolute value $g$ of the dominant eigenvalue of the mixed transfer matrix $\mathcal{T}$,
\begin{equation} \label{eqn:mixedTM}
\mathcal{T} \ \equiv \ \ingr{1.694947}{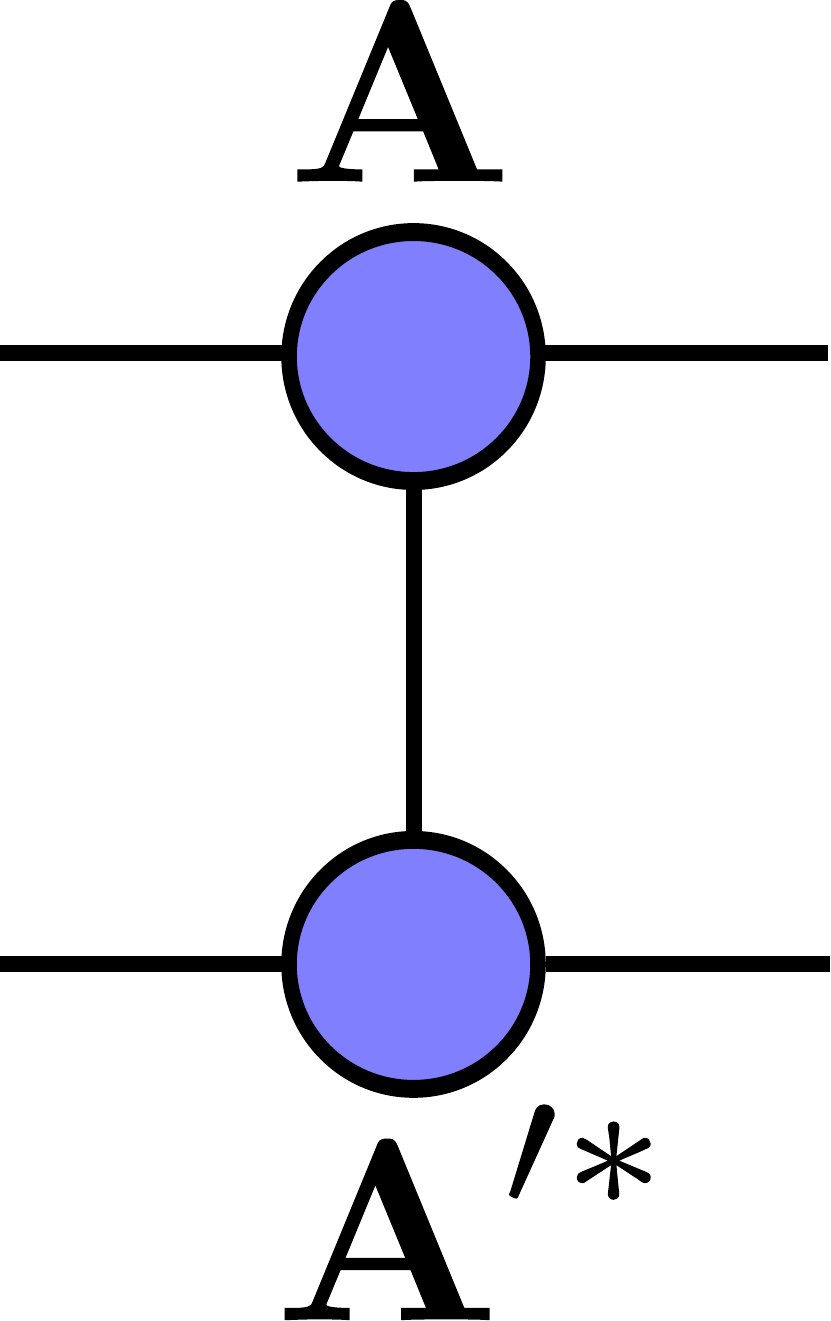} \ .
\end{equation}
For normalized $\ket{\Theta(\lambda)}$ and $\ket{\Theta'(\lambda')}$, we have $g \in [0,1]$. In the case of $g=1$, tensors $\bf A$ and $\bf A'$ describe the same state and are related by a unitary gauge transformation given by the dominant eigenvector of $\mathcal{T}$. Note that the corresponding transfer matrix built by tensors $\bf B$ and $\bf B'$ has the same spectrum, since tensors $\bf A$ and $\bf B$ as well as $\bf A'$ and $\bf B'$ are related by ${\bf B} = \lambda^{-1} {\bf A} \lambda$ and ${\bf B'} = (\lambda')^{-1} {\bf A'} \lambda'$.

For the purpose of this paper we propose a measure of the overlap between localized excitations of the Hamiltonian (\ref{eqn:H}) for nearby values of the parameter $\theta$. In this approach we assess the overlap locally, in the region surrounding the excitations, and neglect the fact that far away from the excitation, states $\ket{\Theta(G)}$ and $\ket{\Theta'(G')}$ have overlap per unit cell strictly smaller than $1$.

To this end, we assume that the ground states $\ket{\Theta(\lambda)}$ and $\ket{\Theta'(\lambda')}$ have the overlap per unit cell close to $1$, which means that tensors $\bf A$ and $\bf A'$ are (up to a gauge) almost identical. This is the case of the ground states of the Hamiltonian (\ref{eqn:H}) for nearby values of the parameter $\theta$, as found in \cite{bauer2014} and further discussed in Section~\ref{sct:results}. In that setting, the relevant overlap between states $\ket{\Theta(G)}$ and $\ket{\Theta'(G')}$ is entirely encoded in their bond matrices $G$ and $G'$ alone. Accordingly, we define the overlap $\braket{\Theta'(G')}{\Theta(G)}$ by comparing $G$ and $G'$ in the scalar product that takes into account the gauge difference between $\bf A$ and $\bf A'$:
\begin{equation}
\braket{\Theta'(G')}{\Theta(G)} \ = \ \ingr{1.719489}{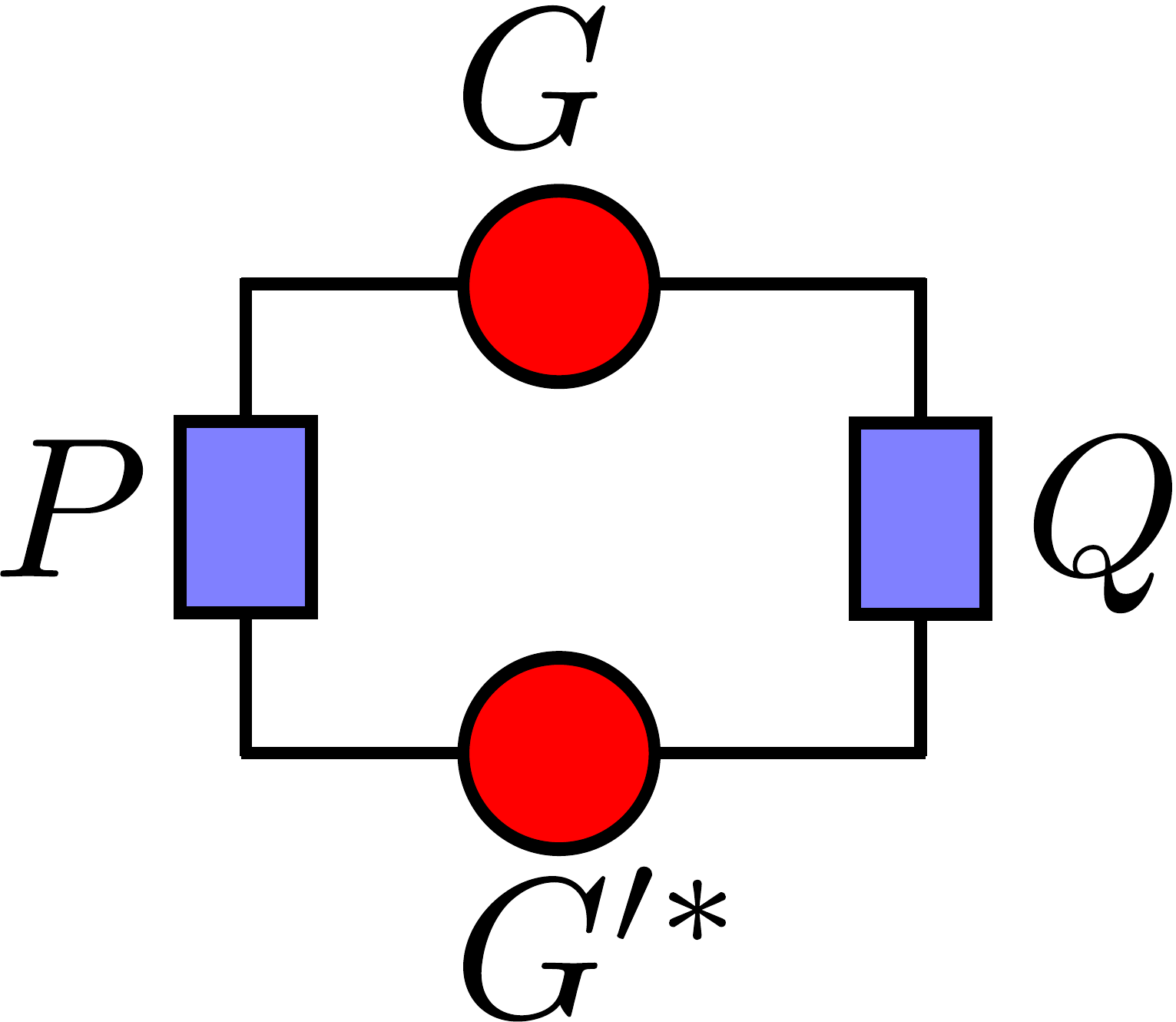} \ ,
\end{equation}
where $P$ is the left dominant eigenvector of the mixed transfer matrix $\mathcal{T}$ defined in \eqnref{eqn:mixedTM}. Similarly, $Q$ is the right dominant eigenvector of the mixed transfer matrix built by tensors $\bf B$ and $\bf B'$. $P$ and $Q$ are normalized such that $\braket{\Theta'(\lambda')}{\Theta(\lambda)}~=~1$.

\bibliography{cslt}

\end{document}